\documentclass[11pt]{article}

\usepackage{amssymb,amsfonts,amsthm}
\usepackage{amsmath}
\usepackage{bm}             
\usepackage[mathscr]{eucal}
\usepackage{cancel}
\usepackage{wasysym}

\usepackage{graphicx}
\usepackage{subfigure}
\def\be{\begin{equation}}
\def\ee{\end{equation}}
\def\bea{\begin{eqnarray}}
\def\eea{\end{eqnarray}}

\def\hsp5{\hspace{5mm}}

\theoremstyle{remark}

\title{\sc Quintessence}

\begin{document}

\author{
\sc Artur Alho,$^{1}$\thanks{Electronic address:{\tt
aalho@math.ist.utl.pt}}\, Elsa Bernholm,$^{2}$\thanks{Electronic address:{\tt
elsabernholm@gmail.com}}\, and Claes Uggla,$^{2}$\thanks{Electronic address:{\tt
claes.uggla@kau.se}}\\
$^{1}${\small\em Center for Mathematical Analysis, Geometry and Dynamical Systems,}\\
{\small\em Instituto Superior T\'ecnico, Universidade de Lisboa,}\\
{\small\em Av. Rovisco Pais, 1049-001 Lisboa, Portugal.}\\
$^{2}${\small\em Department of Physics, Karlstad University,}\\
{\small\em S-65188 Karlstad, Sweden.}}


\date{}
\maketitle

\begin{abstract}

Recent observations suggest that the accelerated expansion of the Universe
at late times is caused by a temporally changing dark energy component, rather than
the constant one in the standard $\Lambda$CDM scenario. In this context quintessence,
i.e. a canonical scalar field minimally coupled to gravity, plays a
prominent role. There are, however, three main types of quintessence models:
thawing quintessence, scaling freezing quintessence, and tracking quintessence.
Dynamical systems reformulations of the field equations
for a broad set of scalar field potentials, including some new ones,
allow us to use dynamical systems methods
to derive global and asymptotic features, visualised in
bounded state space pictures clearly illustrating the relationships and properties
of the different types of quintessence, clarifying initial data issues,
and yielding simple and accurate approximations.

\vspace*{5.0mm}

\emph{Keywords}: General Relativity, Quintessence, dynamical systems.


\end{abstract}

\section{Introduction\label{sec:intro}}

Within the interpretative context of General Relativity (GR), increasingly precise cosmological observations suggest that
(i) the Universe is almost spatially homogeneous, isotropic and flat on large spatial
scales, (ii) it undergoes a late phase of accelerated expansion requiring
a `dark energy' (DE) density component, $\rho_\mathrm{DE}>0$, with a pressure
$p_\mathrm{DE}< -\rho_\mathrm{DE}/3$, (iii) most of the matter in the Universe
is invisible `cold dark matter' (CDM) without pressure.

The simplest model that is compatible with these features is the $\Lambda$CDM
model, for which $\rho_\mathrm{DE} = - p_\mathrm{DE} = \Lambda >0$,
where $\Lambda$ is the cosmological constant. However, $\rho_\mathrm{DE}$ might be dynamical, as suggested by some recent cosmological observations. One of the simplest and theoretically most appealing
dynamical DE models is a canonical scalar field, $\varphi$, minimally coupled to
gravity, referred to as quintessence, the fifth element of the current
matter content in the Universe after radiation, neutrinos, baryons and dark
matter. 

Different spatially homogeneous, isotropic and flat GR quintessence models,
denoted by $\varphi$CDM, are characterised by the scalar field potential
$V(\varphi)$ and the evolution of the equation of state parameter
$w_\mathrm{DE} = w_\varphi \equiv p_\varphi/\rho_\varphi$. For brevity and
simplicity we in this paper only consider monotonically decreasing
potentials $V(\varphi)>0$,
\begin{equation}
\lambda \equiv - \frac{d\ln V}{d\varphi} > 0,
\end{equation}
and matter consisting of baryons and CDM, i.e., matter is modelled as
a pressure-free fluid with energy density $\rho_\mathrm{m} > 0$. Most $\varphi$CDM solutions
are not cosmologically relevant. However, certain particular \emph{quintessence
`attractor' solutions} are of cosmological interest; moreover, they attract
and thereby approximate open sets of other viable solutions, from the matter-dominated
epoch to the asymptotic future.

Quintessence evolution is conveniently described by thawing,
$w_\varphi \approx -1$, $dw_\varphi/da>0$, and freezing,
$w_\varphi >-1$, $dw_\varphi/da<0$, where $a$ is the cosmological scale
factor~\cite{callin05}. There are, however, quintessence models that go through
several periods of both thawing and freezing. To obtain an unambiguous nomenclature we
consider the past asymptotic features of $w_\varphi$ and $\lambda(\varphi)$ for
the attractor solutions, which lead to three main types of quintessence:
\begin{itemize}
\item[(I)] \emph{All} $\varphi$CDM models admit \emph{thawing quintessence} attractor
solutions, parameterised by $\lim_{a\rightarrow 0}\varphi=\varphi_*$ (and thereby
$\lim_{a\rightarrow 0}\lambda = \lambda(\varphi_*) \equiv \lambda_*$) and
$w_\infty \equiv \lim_{a\rightarrow 0}w_\varphi = -1$.
\item[(II)] Models with $\lambda_- = \lim_{a\rightarrow 0}\lambda \gg 1$
and $w_\infty = 0$ admit a single \emph{scaling freezing quintessence} attractor
solution with an initial matter-dominated epoch, $\lim_{a\rightarrow 0}(\rho_\varphi/\rho_\mathrm{m}) \ll 1$.
%
\item[(III)] Models that satisfy $\lim_{\varphi\rightarrow 0}\varphi^{1+r}\lambda(\varphi) = p > 0$,
$r\geq 0$ ($r=0$ ($r>0$) corresponds to asymptotically (exponentially) inverse power-law
potentials), admit a single \emph{tracking quintessence} attractor solution with
$- 1 < w_\infty = -2/(2+p) < 0$ when $r=0$ and $w_\infty =0$ when
$r>0$.\footnote{The name tracking quintessence is also sometimes used in the
literature for scaling freezing quintessence, due to that in both cases there is an
open set of solutions that track, i.e. \emph{shadow}, a single `attractor' solution.}
\end{itemize}

The field equations for the $\varphi$CDM models are the Friedmann and Raychaudhuri
equations, the non-linear Klein-Gordon equation, and the matter conservation equation:
\begin{subequations}\label{Mainsysdim}
\begin{align}
3H^2 &= \rho, \label{Gauss}\\
\dot{H} + H^2 &= -\frac16(\rho + 3p), \label{Ray}\\
\ddot{\varphi} &=-3H\dot{\varphi} - V_{,\varphi}, \label{KG}\\
\dot{\rho}_\mathrm{m} &= -3H\rho_\mathrm{m},\label{meq}
\end{align}
\end{subequations}
where an overdot represents the derivative with respect to the cosmic
proper time $t$ whereas the Hubble scalar is defined by $H=\dot{a}/a$;
the total energy density $\rho$ and pressure $p$ are given by
\begin{equation}\label{rhoptot}
\rho = \rho_\varphi + \rho_\mathrm{m},\qquad
p = p_\varphi,
\end{equation}
where
\begin{equation}\label{rhophipphi}
\rho_\varphi = \frac12\dot{\varphi}^2 + V(\varphi),\qquad
p_\varphi = \frac12\dot{\varphi}^2 - V(\varphi).
\end{equation}
Since $\rho>0$ it follows from~\eqref{Gauss} that $H(t)>0$ for initially
expanding models. By introducing $\dot{\varphi}$ as a variable the above equations
form a four-dimensional (4D) dynamical system\footnote{A (continuous) dynamical
system is a set of autonomous first order ordinary differential equations
$\dot{{\bf x}} = {\bf f}({\bf x})$, ${\bf x} \subseteq \mathbb{R}^n$. A regular
dynamical system requires that ${\bf f}({\bf x})$ is at least $C^1$-differentiable,
which enables a local eigenvalue analysis at fixed points (also known
as critical points, equilibrium points) ${\bf x}_0$, ${\bf f}({\bf x}_0)=0$.}
for the state vector $(H,\varphi,\dot{\varphi},\rho_\mathrm{m})$ obeying the
constraint~\eqref{Gauss}.

In the next two sections we introduce \emph{dimensionless} variables
and 3D \emph{unconstrained regular} dynamical systems, yielding
state space pictures that show that the scaling freezing and tracking quintessence
attractor solutions, respectively, form the interior state space boundary of one-parameter
sets of thawing quintessence solutions.
Section~\ref{sec:approx} introduces quintessence approximations that
are simpler and more accurate than earlier ones in the literature. Finally,
Section~\ref{sec:disc} discusses dynamical systems formulations
and features for generalisations of the present illustrative examples and for
other types of scalar field potentials, and it also provides
references outlining the history of scalar field cosmology.


\section{Thawing and scaling freezing quintessence\label{sec:scale}}

We begin by introducing some physically interesting dimensionless Hubble-normalised
quantities: the Hubble-normalised matter and scalar field energy densities, where the
latter consists of the Hubble-normalised kinetic and potential parts,
\begin{equation}
\Omega_\mathrm{m} \equiv \frac{\rho_\mathrm{m}}{3H^2},\qquad \Omega_\varphi \equiv \frac{\rho_\varphi}{3H^2} = \Omega_T + \Omega_V, \qquad
\Omega_T \equiv \frac{\dot{\varphi}^2}{6H^2},\qquad \Omega_V \equiv \frac{V(\varphi)}{3H^2}.
\end{equation}
It follows that~\eqref{Gauss} then yields
\begin{equation}\label{constrom}
\Omega_\varphi + \Omega_\mathrm{m} = \Omega_T + \Omega_V + \Omega_\mathrm{m} = 1,
\end{equation}
wheras~\eqref{rhophipphi} results in
\begin{equation}\label{wphidef}
w_\varphi \equiv \frac{p_\varphi}{\rho_\varphi} = \frac{\Omega_T - \Omega_V}{\Omega_T + \Omega_V} = -1 + \frac{2\Omega_T}{\Omega_T + \Omega_V} \in [-1,1],
\end{equation}
where $w_\varphi = - 1$ when $\dot{\varphi} = 0$, $\Omega_T=0$, $\Omega_V>0$,
while $w_\varphi = 1$ when $V=0$, $\Omega_V=0$, $\Omega_T>0$.
As follows from~\eqref{Gauss} and~\eqref{Ray}, the \emph{deceleration parameter} $q$ is given by
\begin{equation}
q \equiv -\frac{a\ddot{a}}{\dot{a}^2} = -\left(1 + \frac{H^\prime}{H}\right) = \frac12\left(1 + 3w_\varphi\Omega_\varphi\right)
= \frac12\left[1 + 3\left(\Omega_T - \Omega_V\right)\right] \in [-1,2],
\end{equation}
where a ${}^\prime$ henceforth denotes 
the derivative with respect to the $e$-fold time
\begin{equation}
N \equiv \ln(a/a_0),\qquad a_0 = a(t_0),
\end{equation}
where $t_0$ is some reference time, from which it
follows that $dN/dt = H$ and $a \rightarrow 0\, \Rightarrow\, N \rightarrow -\infty$;
$q=-1$ when $w_\varphi = -1$ and $\Omega_T=0$, $\Omega_V = 1$ while $q=2$ when $w_\varphi = 1$
and $\Omega_T = 1$, $\Omega_V=0$.

Next we follow~\cite{alhetal23} (AUW1) and introduce
\begin{equation}
u \equiv \frac{\dot{\varphi}}{\sqrt{\rho_\varphi}} = \frac{\varphi^\prime}{\sqrt{3\Omega_\varphi}},\qquad
v \equiv \sqrt{\frac{\Omega_\varphi}{3}},\label{uvdef}
\end{equation}
which leads to
\begin{subequations}\label{uvinterpret}
\begin{alignat}{3}
w_\varphi &= u^2 - 1,&\qquad \Omega_\mathrm{m} &= 1 - 3v^2,&\qquad \Omega_\varphi &= 3v^2,\\
\Omega_T &= \frac32\left(uv\right)^2,&\qquad \Omega_V &= \frac32\left(2 - u^2\right)v^2,&\qquad
q &= \frac12\left(1 + 9(u^2 - 1)v^2\right),
\end{alignat}
\end{subequations}
and thus
\begin{equation}
u\in[-\sqrt{2},\sqrt{2}],\qquad v\in[0,1/\sqrt{3}],
\end{equation}
where $u=\mathrm{const.}\,\Rightarrow\, w_\varphi = \mathrm{const.}$;
$v=\mathrm{const.}\,\Rightarrow\, \Omega_\varphi = \mathrm{const.}$,
with boundary values $u = \pm\sqrt{2}\,\Rightarrow\, w_\varphi = 1,\, \Omega_V = 0$;
$v=0\,\Rightarrow\,\Omega_\varphi=0,\, \Omega_\mathrm{m}= 1$;
$v = 1/\sqrt{3}\,\Rightarrow\,\Omega_\varphi=1,\,\Omega_\mathrm{m}=0$.

The variables $u$ and $v$ are appropriate for models with bounded
$\lambda(\varphi)$ and
$\lim_{\varphi\rightarrow \pm\infty}\lambda(\varphi) = \lambda_\pm = \mathrm{const.}$,
but in order to obtain a bounded state space, which
is a desirable property both for mathematical and illustrative purposes,
we also need to introduce a bounded variable monotonically increasing in $\varphi$,
$\bar{\varphi}\in [\bar{\varphi}_-,\bar{\varphi}_+]$, where
we for simplicity require
$\bar{\varphi}_\pm \equiv \lim_{\varphi\rightarrow \pm\infty}\bar{\varphi} = \pm 1$.
This leads to the following dynamical system\footnote{Using
$w_\varphi$ and $\Omega_\varphi$ instead of $u$ and $v$ lead to a
non-regular dynamical system.}
\begin{subequations}\label{Dynsys.uv}
\begin{align}
{\bar\varphi}^\prime &= 3uvG(\bar{\varphi}),\label{barvarphiprime}\\
u^\prime &= \frac{3}{2}(2-u^2)(v\lambda(\bar{\varphi})-u), \label{u.prime} \\
v^\prime &= \frac{3}{2}(1-u^2)(1-3v^2)v, \label{v.prime}
\end{align}
\end{subequations}
where
\begin{equation}
G(\bar{\varphi}) \equiv \left(\frac{d\bar{\varphi}}{d\varphi}\right)> 0,\quad \bar{\varphi} \in (-1,1);\qquad G(\pm 1) = 0.
\end{equation}

For a wide class of potentials $V(\varphi)$, $\bar{\varphi}(\varphi)$ can be adapted to the
properties of $\lambda(\varphi)$ so that both $G(\bar{\varphi})$ and, with a slight abuse
of notation, $\lambda(\bar{\varphi})$ are regular functions and $\lambda(\bar{\varphi})$
has limits $\lambda_\pm = \lambda(\pm 1)$, $\lambda_- \geq \lambda_+ \geq 0$, where
$\lambda_+=\lambda_-\,\Rightarrow\, \lambda = \mathrm{constant} = \lambda_+=\lambda_-$,
which yields $V = V_*\exp(-\lambda\varphi)$. Eq.~\eqref{Dynsys.uv} then forms an
\emph{unconstrained regular} dynamical system on a 3D \emph{`box state space'}
$(\bar{\varphi},u,v)$ bounded by 2D invariant boundary sets.

\begin{figure}[ht!]
	\begin{center}
		\subfigure[$\lambda=0$]{\label{fig:LCDM}
			\includegraphics[width=0.27\textwidth]{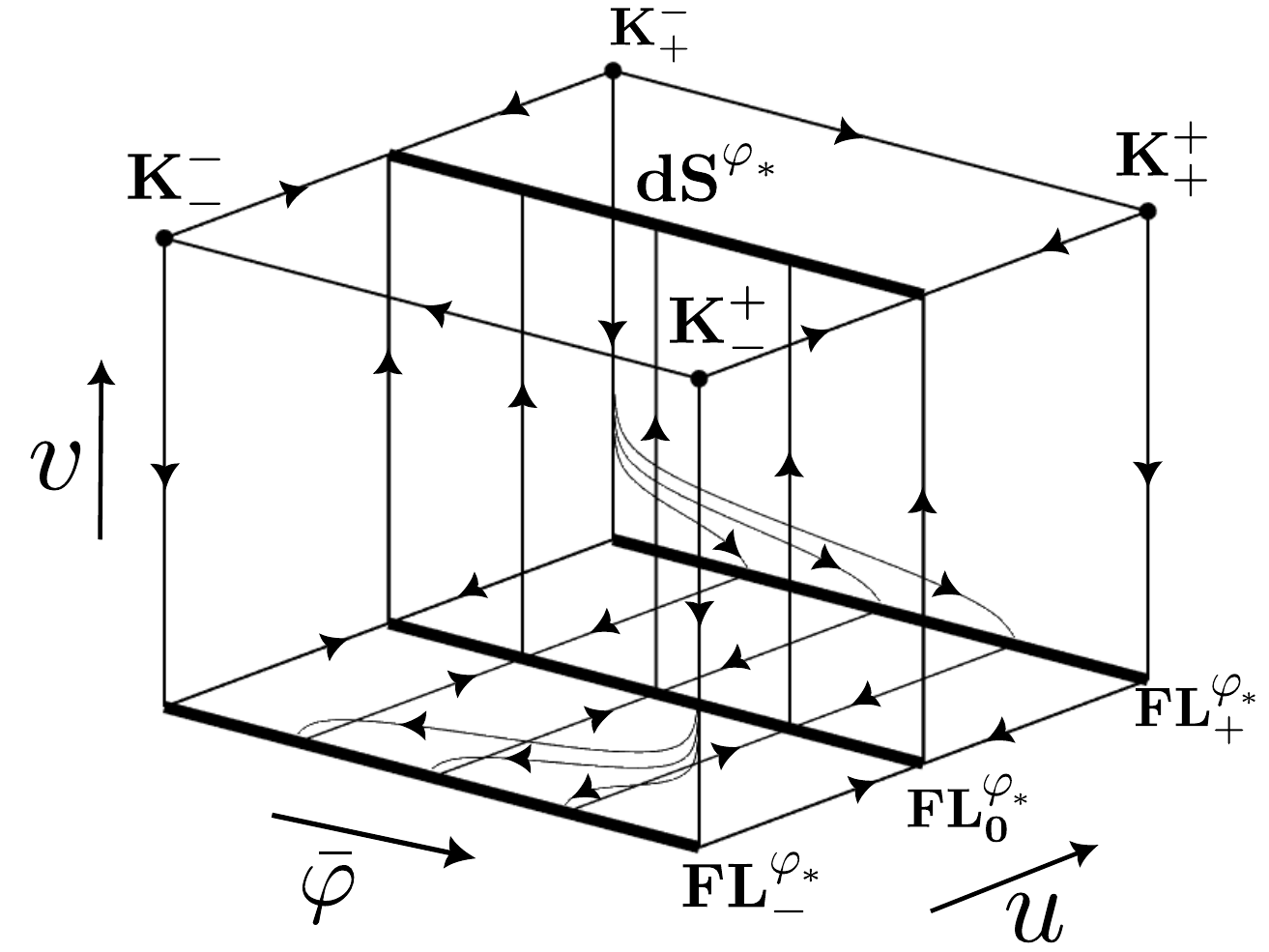}}
		\subfigure[$\lambda=1$]{\label{fig:deform}
			\includegraphics[width=0.27\textwidth]{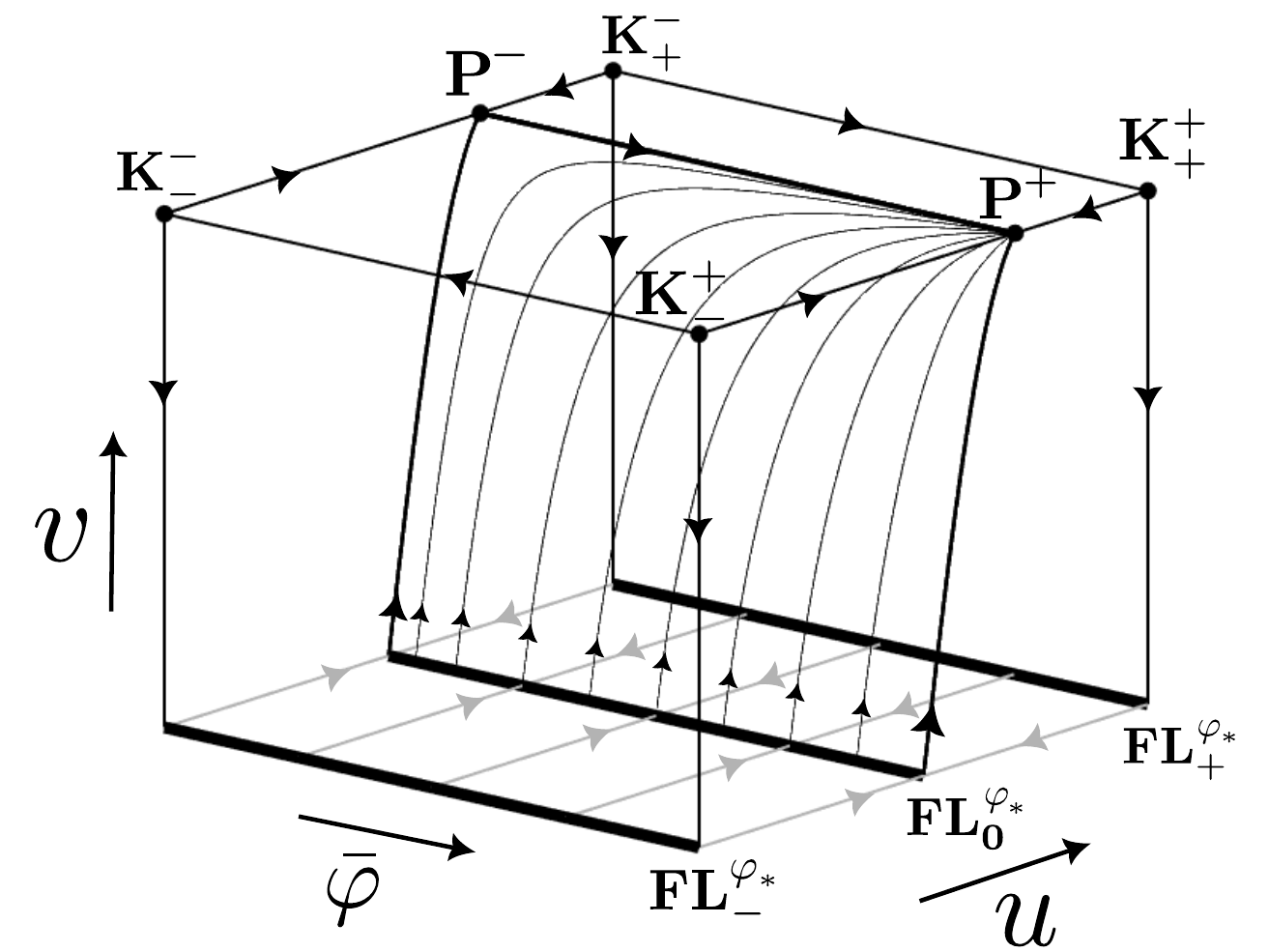}}
		\subfigure[$\lambda=10$]{\label{fig:scaling}
			\includegraphics[width=0.27\textwidth]{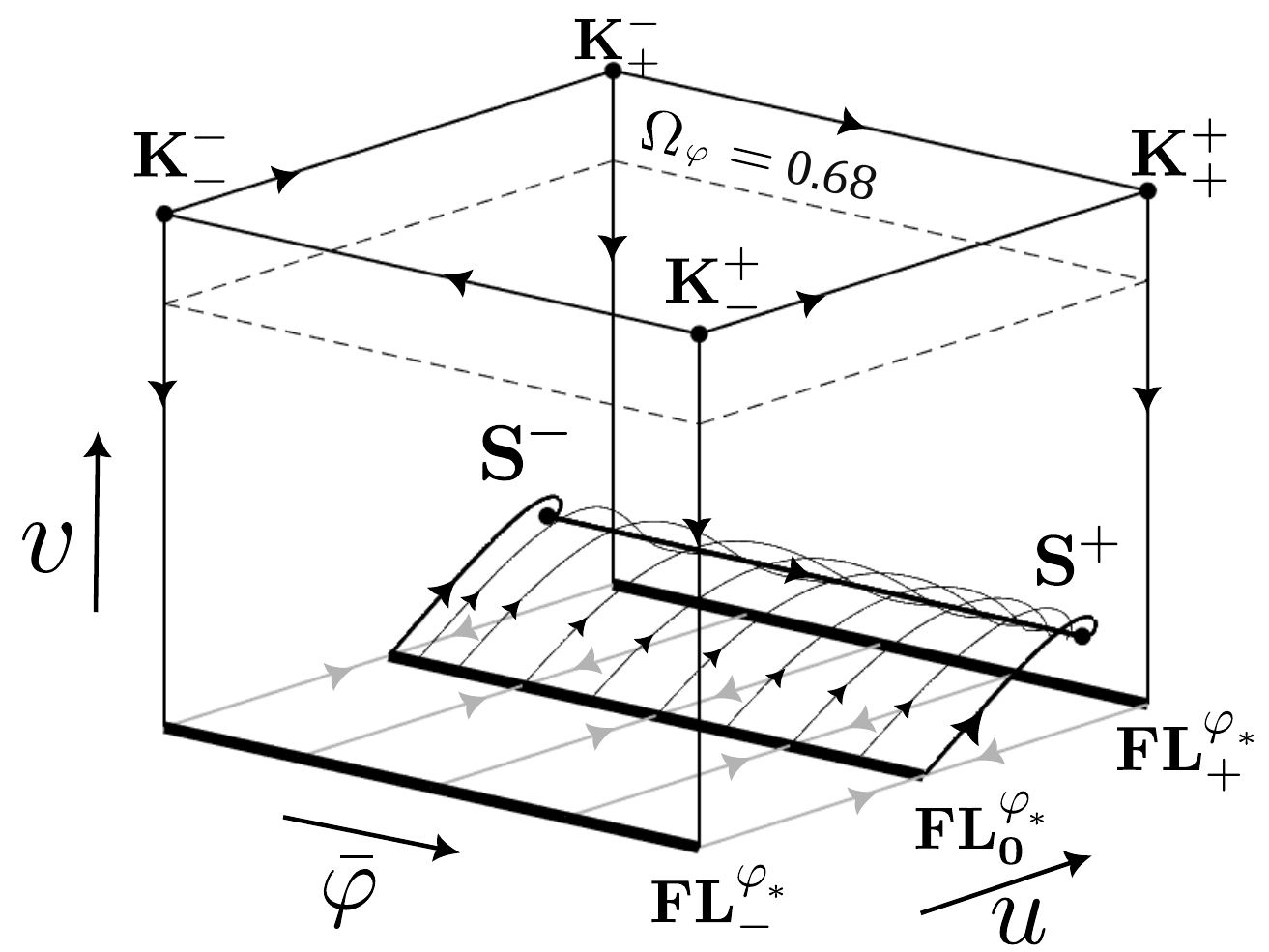}}
		\vspace{-0.5cm}
	\end{center}
	\caption{Solutions $\mathrm{FL}_\pm^{\varphi_*} \rightarrow \mathrm{FL}_0^{\varphi_*}$
on the matter-dominated boundary $v=0$ ($\Omega_\mathrm{m}=1$),
which provide the thawing quintessence attractor mechanism,
and solutions on the unstable $\mathrm{FL}_0^{\varphi_*}$ manifold surface
(${\bf U}\mathrm{FL}_0^{\varphi_*}$) for $V \propto \exp(-\lambda\varphi)$.
Figure (a) also depicts solutions
$\mathrm{K}_\pm^\mp \rightarrow \mathrm{FL}_\pm^{\varphi_*}$
on the boundaries $u=\pm \sqrt{2}$
($\Omega_V=0$). Figure (b) illustrates that there are classes of potentials that
continuously deform the $\Lambda$CDM ${\bf U}\mathrm{FL}_0^{\varphi_*}$ with $u=0$ in Figure (a)
to a thawing quintessence ${\bf U}\mathrm{FL}_0^{\varphi_*}$.
The boundary $\mathrm{P}^-\rightarrow\mathrm{P}^+$ of ${\bf U}\mathrm{FL}_0^{\varphi_*}$
in Figure (b) is replaced with the scaling solution $\mathrm{S}^-\rightarrow\mathrm{S}^+$
in Figure (c) when $\lambda >\sqrt{3}$,
where the scaling matter-dominant condition $\lambda \gg 1$ implies that $\Omega_\varphi$ never
becomes observationally significant for either type of quintessence.
}
\label{fig:constlambda}
\end{figure}

The equations are easily solved on the boundaries $v=0$ and $u=\pm\sqrt{2}$
where the solutions are independent of $\lambda(\bar{\varphi})$ and $V(\varphi)$: On the 
matter-dominated Friedmann-Lema\^{i}tre boundary $v=0$ ($\Omega_\mathrm{m}=1$) there are 
three lines of fixed points, $\mathrm{FL}^{\varphi_*}_0$, with $u=0$ ($w_\varphi = -1$) 
and $\mathrm{FL}^{\varphi_*}_\pm$, with $u =\pm\sqrt{2}$ ($w_\varphi = 1$), connected by 
the solutions $\mathrm{FL}^{\varphi_*}_\pm \rightarrow \mathrm{FL}^{\varphi_*}_0$ with
$\bar{\varphi} = \bar{\varphi}_* = \mathrm{const.}$; on $u = \pm \sqrt{2}$ the
solutions originate from the `kinaton' fixed points $\mathrm{K}_\pm^\mp$
($\Omega_T = 1$) where each solution end at a distinct fixed point on
$\mathrm{FL}^{\varphi_*}_\pm$, see Figure~\ref{fig:LCDM}.

It follows from~\eqref{barvarphiprime} that $\bar{\varphi}(N) \in (-1,1)$ is
monotonically increasing (decreasing) when $v>0$ and $u>0$ ($u<0$). For
monotonically decreasing potentials, $\lambda(\bar{\varphi}) >0$,
$\bar{\varphi} \in (-1,1)$, it follows from~\eqref{u.prime} that all
`interior' solutions reach the region $u>0$, since $u$ is monotonically increasing
when $v>0$ and $u \leq 0$. Moreover, $3H^2 = 2V(\bar{\varphi})/3(2-u^2)v^2$
is a monotonic function, which prevents fixed points and recurring orbits
when $v>0$ and $\bar{\varphi} \in (-1,1)$, and hence all such solutions
originate and end at the boundaries $v=0$ or
$\bar{\varphi} = \pm 1$. In addition, the equations on
$\bar{\varphi} = \pm 1$ form the same coupled system of equations for $(u,v)$ as
when $V = V_*\exp(-\lambda_\pm\varphi)$.
When all this is taken together with the known solution structure on the
boundaries, it follows that \emph{all solutions} for the present
class of models \emph{originate and end at different fixed points}
residing on the boundaries $v=0$ and $\bar{\varphi} = \pm 1$.

\emph{The line of fixed points} $\mathrm{FL}^{\varphi_*}_0$
\emph{is central for thawing quintessence}.
Linearisation of~\eqref{Dynsys.uv} at $\mathrm{FL}_0^{\varphi_*}$,
where $w_\varphi=-1$, results in that at each fixed point on
$\mathrm{FL}_0^{\varphi_*}$ there is
\begin{itemize}
\item[(i)] a zero eigenvalue $\Rightarrow$ the line of fixed points $\mathrm{FL}_0^{\varphi_*}$;
\item[(ii)] a negative eigenvalue $\Rightarrow$ the solutions
$\mathrm{FL}^{\varphi_*}_\pm \rightarrow \mathrm{FL}^{\varphi_*}_0$ on $v=0$
($\Omega_\mathrm{m} = 1$), with `frozen' scalar field values
$\bar{\varphi} = \bar{\varphi}_* = \mathrm{const.}$, constituting the stable manifold
of $\mathrm{FL}_0^{\varphi_*}$;
\item[(iii)] a positive eigenvalue $\Rightarrow$ a single solution originating
from each fixed point on $\mathrm{FL}_0^{\varphi_*}$ with an eigenvector with $u>0$,
except for the $\Lambda$CDM case with $\lambda=0$, see the Figures below. Taken
together, these solutions form the 2D unstable manifold surface of solutions of
$\mathrm{FL}_0^{\varphi_*}$.
\end{itemize}

\emph{The observational requirement of a long matter-dominated epoch},
$\Omega_\mathrm{m} \approx 1$ ($v\approx 0$), \emph{poses severe restrictions on viable
initial data}. For monotonically decreasing potentials generic solutions
originate from fixed points on the \emph{kinaton boundary} $\Omega_T=1$.
It is only a `small' open set of these solutions that is relevant
for thawing quintessence, since a long matter-dominated epoch requires
thawing quintessence solutions to come very close to the matter-dominated
$v=0$ boundary. There these solution closely shadow the
solutions $\mathrm{FL}^{\varphi_*}_\pm \rightarrow \mathrm{FL}^{\varphi_*}_0$
and are thereby pushed to the `attractor' solutions on the unstable manifold of
$\mathrm{FL}^{\varphi_*}_0$, which hence approximate them extremely well
during their observationally relevant evolution --- \emph{this is the thawing
quintessence `attractor' mechanism} (together with, as we will see, the existence
of a common fixed point as the sink and true future attractor). Finally,
note that $\varphi$ acts as a frozen test field that does not affect the spacetime
geometry during the matter-dominated epoch, where $u(N)$ (and thereby $w_\varphi(N)$)
does not describe anything observable.

As an illustration, consider first
$V = V_*\exp(-\lambda\varphi)$ 
for which the equation for $\bar{\varphi}$ decouples, resulting in a 2D reduced
dynamical system of coupled equations for $u$ and $v$.
However, we also consider $\bar{\varphi}$ in order to make comparisons since all 
other potentials yield non-trivial $\lambda(\bar{\varphi})$ and 3D
coupled systems of equations; a simple choice of $\bar{\varphi}$ is
$\bar{\varphi} \equiv \tanh(\varphi)\,\Rightarrow G(\bar{\varphi}) = 1 - \bar{\varphi}^2,\,\bar{\varphi}\in [-1,1]$.
Due to the decoupling of $\bar{\varphi}$, projecting solutions with
$\bar{\varphi}\in (-1,1)$ onto the $(u,v)$-plane results in the same
trajectories as those on the boundaries $\bar{\varphi}=\pm 1$.

When $V=\Lambda>0\,\Rightarrow\,\lambda = 0$, the unstable manifold of
$\mathrm{FL}^{\varphi_*}_0$, referred to as
${\bf U}\mathrm{FL}_0^{\varphi_*}$, is given by the $\Lambda$CDM solutions
$\mathrm{FL}^{\varphi_*}_0 \rightarrow \mathrm{dS}^{\varphi_*}$
on the, in this case, invariant $u=0$ subset, see eq.~\eqref{u.prime},
with $\bar{\varphi} = \bar{\varphi}_* = \mathrm{const.}$, where
$\mathrm{dS}$ stands for de Sitter. Increasing $\lambda$
results in increasing deformations of the ${\bf U}\mathrm{FL}_0^{\varphi_*}$
surface of solutions. When $\lambda>\sqrt{3}$ the solution
$\mathrm{S}^-\rightarrow \mathrm{S}^+$, with $(u,v) = (1,1/\lambda)$ and
thereby $w_\varphi = 0$, $\Omega_\varphi = 3/\lambda^2$, replaces $\mathrm{P}^-\rightarrow \mathrm{P}^+$, 
with $(u,v) = (\lambda,1)/\sqrt{3}$, as the $\bar{\varphi}\in (-1,1)$-boundary
of the ${\bf U}\mathrm{FL}_0^{\varphi_*}$ surface of solutions
(where $\mathrm{P}^+$ ($\mathrm{S}^+$)
is the future attractor when $\lambda \leq \sqrt{3}$
($\lambda > \sqrt{3}$)).\footnote{The kernel $\mathrm{P}$ stands
for Power-law since $0 < \lambda < \sqrt{2}$ yields power-law accelerated
expansion. The kernel $\mathrm{S}$ denotes Scaling, since
$v = 1/\lambda \Rightarrow \rho_\varphi(N)/\rho_\mathrm{m}(N) =
\Omega_\varphi/\Omega_\mathrm{m} = \Omega_\varphi/(1 - \Omega_\varphi)= 3v^2/(1-3v^2) =
3/(\lambda^2-3)$, and hence $\rho_\mathrm{m}(N) \propto \exp(-3N) \propto a^{-3}$ and
$\rho_\varphi(N)$ `scale' in the same manner (are proportional).}
See Figure~\ref{fig:constlambda} and AUW1 for further details.

Next we consider \emph{scaling freezing quintessence}, which requires potentials
with varying $\lambda(\bar{\varphi})$, where early matter-domination
and future eternal accelerated expansion demands $\lambda_- \gg 1$,
$0 \leq \lambda_+  \lessapprox {\cal O}(1)$, respectively.
Following AUW1, we illustrate the relationship between
scaling freezing and thawing quintessence with the potential
\begin{equation} \label{two.exp.pot}
V = M_-^4e^{-\lambda_-\varphi}+ M_+^4e^{-\lambda_+\varphi},\qquad
M_\pm >0,\qquad \lambda_- > \lambda_+ \geq 0,
\end{equation}
where we choose
\begin{equation}
\bar{\varphi} = \tanh(C\varphi + D), \qquad C = \frac12(\lambda_- -\lambda_+),\qquad
D = \ln[(M_+/M_-)^4],
\end{equation}
which leads to
\begin{equation} \label{lambda.bar.phi}
\lambda(\bar{\varphi}) = \frac12\lambda_+(1 + \bar{\varphi}) +
\frac12\lambda_-(1 - \bar{\varphi}),\qquad G(\bar{\varphi}) = C(1-\varphi^2).
\end{equation}

As seen in Figure~\ref{fig:scaling}, the thawing quintessence
${\bf U}\mathrm{FL}_0^{\varphi_*}$ surface can be viewed as a continuous deformation
in $u$ and $v$ of its boundary solutions
$\mathrm{FL}_0^{\varphi_*} \rightarrow \mathrm{S}^-$ and
$\mathrm{FL}_0^{\varphi_*} \rightarrow \mathrm{P}^+$
($\mathrm{dS}^+$ if $\lambda_+ =0$) on $\bar{\varphi} = \pm 1$, while the
\emph{scaling freezing solution} $\mathrm{S}^-\rightarrow \mathrm{P}^+$ forms
the interior boundary of ${\bf U}\mathrm{FL}_0^{\varphi_*}$, describing
the late evolution of thawing quintessence. By choosing $\lambda_\pm$ appropriately
both the scaling freezing and thawing solutions become observationally
viable. \emph{The scaling freezing quintessence attractor mechanism} is due to that
the $\bar{\varphi} = - 1$ boundary is the stable manifold of $\mathrm{S}^-$,
while the scaling freezing solution is its unstable manifold. A long matter-dominated
scaling epoch requires initial data to be extremely close to $\mathrm{S}^-$
with $\bar{\varphi} \approx - 1$, where the vicinity to the stable manifold
of $\mathrm{S}^-$ pushes an open set of solutions toward the solution
$\mathrm{S}^-\rightarrow \mathrm{P}^+$ near $\mathrm{S}^-$, where
$\mathrm{S}^-\rightarrow \mathrm{P}^+$ is subsequently shadowed and
thereby approximate these solutions extremely well, a phenomenon further
strengthened by that $\mathrm{P}^+/\mathrm{dS}^+$ is the true future attractor
where all interior solutions end.

The next example is the following inflationary $\alpha$-attractor quintessence potential,
introduced in~\cite{dimowe17} 
and studied 
from a dynamical systems perspective in~\cite{alhugg23}, although we here omit radiation:
\begin{equation}
V = V_*\left(\frac{e^{-\nu(1 + \bar{\varphi})} - e^{-2\nu}}{1 - e^{-2\nu}}\right),\qquad
\bar{\varphi} \equiv \tanh\frac{\varphi}{\sqrt{6\alpha}},\qquad
\lambda = \frac{\nu}{\sqrt{6\alpha}}
\left(\frac{1 - \bar{\varphi}^2}{1 - e^{-\nu(1 - \bar{\varphi})}}\right).
\end{equation}
Inflationary $\alpha$-attractor arguments suggest that $\alpha \sim \mathcal{O}(1)$
while inflationary and quintessence energy considerations result in large values for
$V_*$, $\varphi_*$ and $\nu \sim \mathcal{O}(125)$, see~\cite{akretal18}. We choose
$\alpha = 7/3$ and, for illustrative purposes, $\nu = 15$.

\begin{figure}[ht!]
	\begin{center}
		\subfigure[$\lambda_- = 10$, $\lambda_+ = 1$, $M_+ = M_-$]{\label{fig:scaling}
			\includegraphics[width=0.30\textwidth]{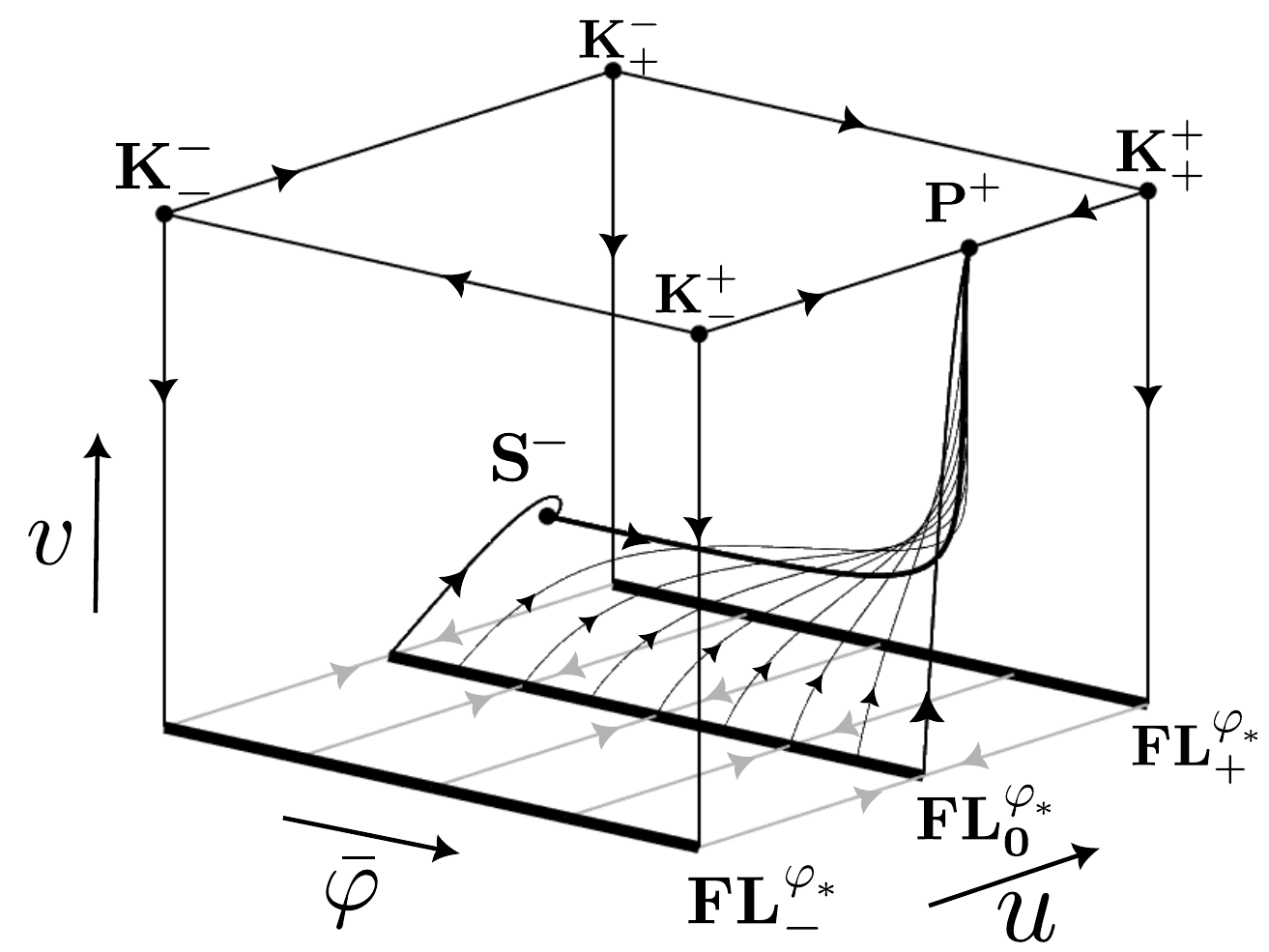}}
		\hspace{1.0cm}
		\subfigure[$\alpha = \frac73$, $\nu = 15$]{\label{fig:alpha_attr}
			\includegraphics[width=0.30\textwidth]{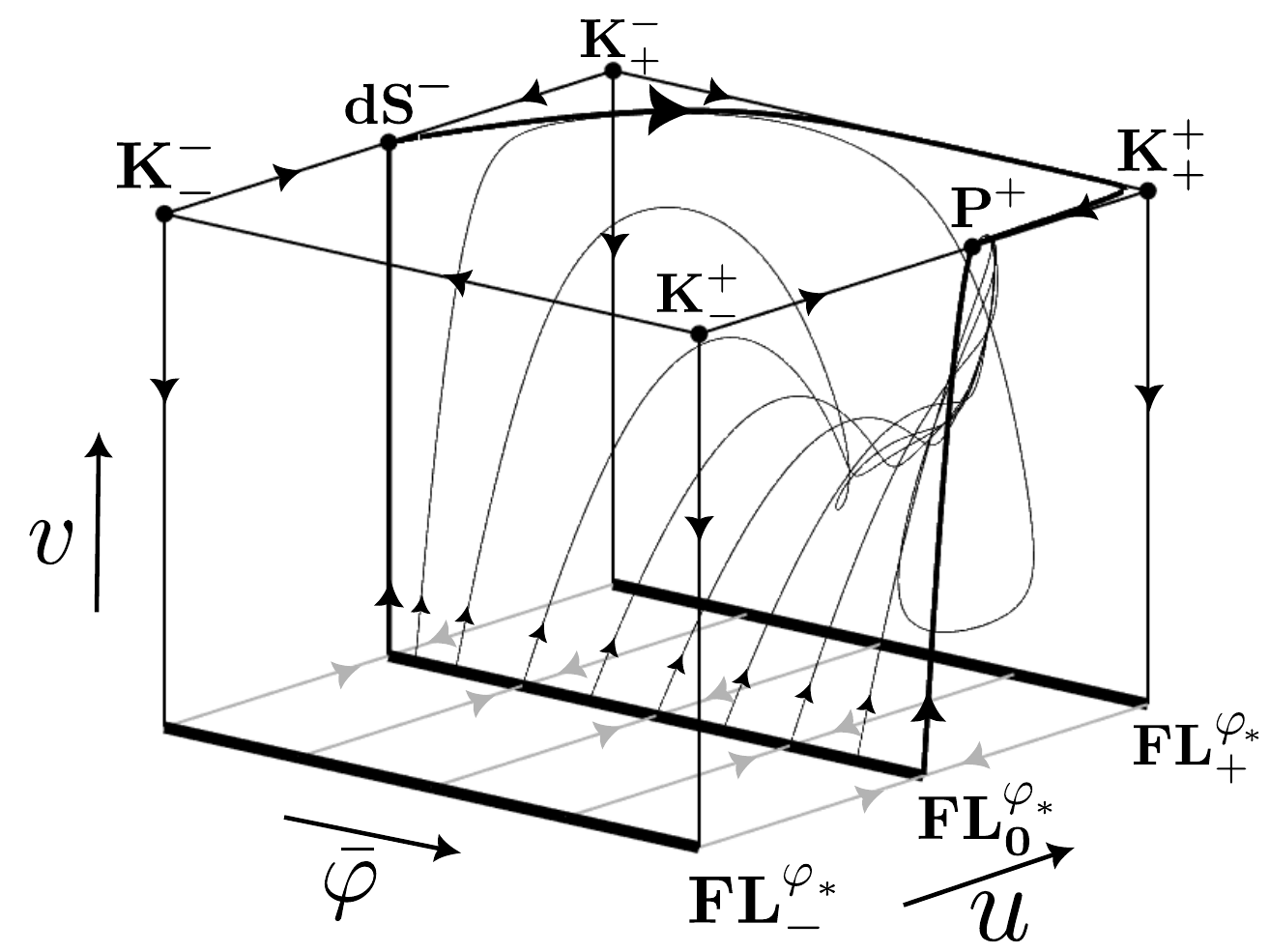}}
		\vspace{-0.5cm}
	\end{center}
	\caption{(a) Depicts the surface of thawing quintessence attractor solutions with the
             scaling freezing attractor solution $\mathrm{S}^- \rightarrow \mathrm{P}^+$ as
             its interior boundary for the potential
             $V = M_-^4e^{-\lambda_-\varphi}+ M_+^4e^{-\lambda_+\varphi}$. (b) Depicts the inflationary
             $\alpha$-attractor solution for
             $V \propto \frac{e^{-\nu(1 + \tanh\frac{\varphi}{\sqrt{6\alpha}})} - e^{-2\nu}}{1 - e^{-2\nu}}$
             on the scalar field boundary $v=1/\sqrt{3}$ ($\Omega_\varphi = 1$)
             and the surface of thawing quintessence attractor solutions, where only those with
             $\bar{\varphi}_*\approx 1$ are cosmologically viable.
}
\label{Fig_scaling_alphaattr}
\end{figure}

Figure~\ref{fig:alpha_attr} depicts the \emph{inflationary attractor solution}
$\mathrm{dS}^-\rightarrow \mathrm{P}^+$ on the scalar
field boundary $v=1/\sqrt{3}$ ($\Omega_\varphi = 1$) and thawing attractor solutions on
${\bf U}\mathrm{FL}_0^{\varphi_*}$. The inflationary attractor solution is the unstable
\emph{centre} manifold of the fixed point $\mathrm{dS}^-$, corresponding to an eigenvalue that
is zero, while $\mathrm{dS}^-$ is a sink \emph{on} the $\bar{\varphi} = -1$ boundary associated
with two negative eigenvalues. This latter feature pushes nearby solutions toward the inflationary
attractor solution where the zero eigenvalue strengthens its attracting nature and
also enables solutions to stay in a long inflationary quasi-de Sitter epoch in the vicinity
of $\mathrm{dS}^-$. Figure~\ref{fig:alpha_attr} also illustrates that the inflationary attractor
solution goes through a 'kinaton' epoch ($\Omega_T \approx 1$) when it shadows the kinaton solution
$\mathrm{K}_+^- \rightarrow \mathrm{K}_+^+$, a feature that is enhanced with increasing $\nu$,
see~\cite{alhugg23}. Note that $v$ ($\Omega_\varphi$) is monotonically decreasing when $u>1$ ($w_\varphi > 0$),
which makes most thawing quintessence solutions observationally non-viable --- it is only those with
$\bar{\varphi}_* \approx 1$ (corresponding to values $\varphi_*$ on the exponential tail of the
potential where $\lambda(\varphi_*) \lessapprox {\cal O}(1)$) that are observationally viable.
%

%
%

\section{Thawing and tracking quintessence\label{sec:track}}

The term tracking quintessence was introduced by Steinhardt et al.
in~\cite{steetal99,zlaetal99} for models with $V\propto \varphi^{-p}$, $p>0$ and
$V \propto \exp(M/\varphi) - 1$. In~\cite{alhetal24} (AUW2) the first case was
generalised to asymptotically inverse power-law potentials when
$\varphi \rightarrow 0$ and here we will also consider generalisations of
the second case. In contrast to what was stated by Tsujikawa in the
review~\cite{tsu13}, there are three types of tracking quintessence:
\emph{tracking freezing, tracking constant and tracking thawing quintessence},
as illustrated by the asymptotically inverse power-law potentials
\begin{equation}\label{tanhpot}
V = V_*\left[\frac{\nu (\cosh\nu\varphi)^{1-\alpha}}{\sinh(\nu\varphi)}\right]^p
\end{equation}
discussed in AUW2. For brevity we here only consider $\alpha=1$, i.e.,
\begin{equation}
V = V_*\sinh^{-p}(\nu\varphi),\quad p>0,\quad \nu >0 \quad \Rightarrow \quad
\lambda = \frac{p\nu}{\tanh{(\nu\varphi)}} > 0.
\end{equation}
It follows that $\lambda_+=p\nu$. We choose
$\bar{\varphi} = \tanh{(\nu\varphi)}$, $\bar{\varphi} \in [0,1]$.
The previous variable $v$ now results in that the equations become
irregular at $\bar{\varphi}=0$ since
$\lim_{\bar{\varphi}\rightarrow 0}(\bar{\varphi}\lambda(\bar{\varphi})) = p$.

\begin{figure}[ht!]
	\begin{center}
		\subfigure[$p = \frac{14}{3}$, $\nu = \frac{3}{14}\sqrt{\frac{3}{10}}$]{\label{fig:track_freeze}
			\includegraphics[width=0.30\textwidth]{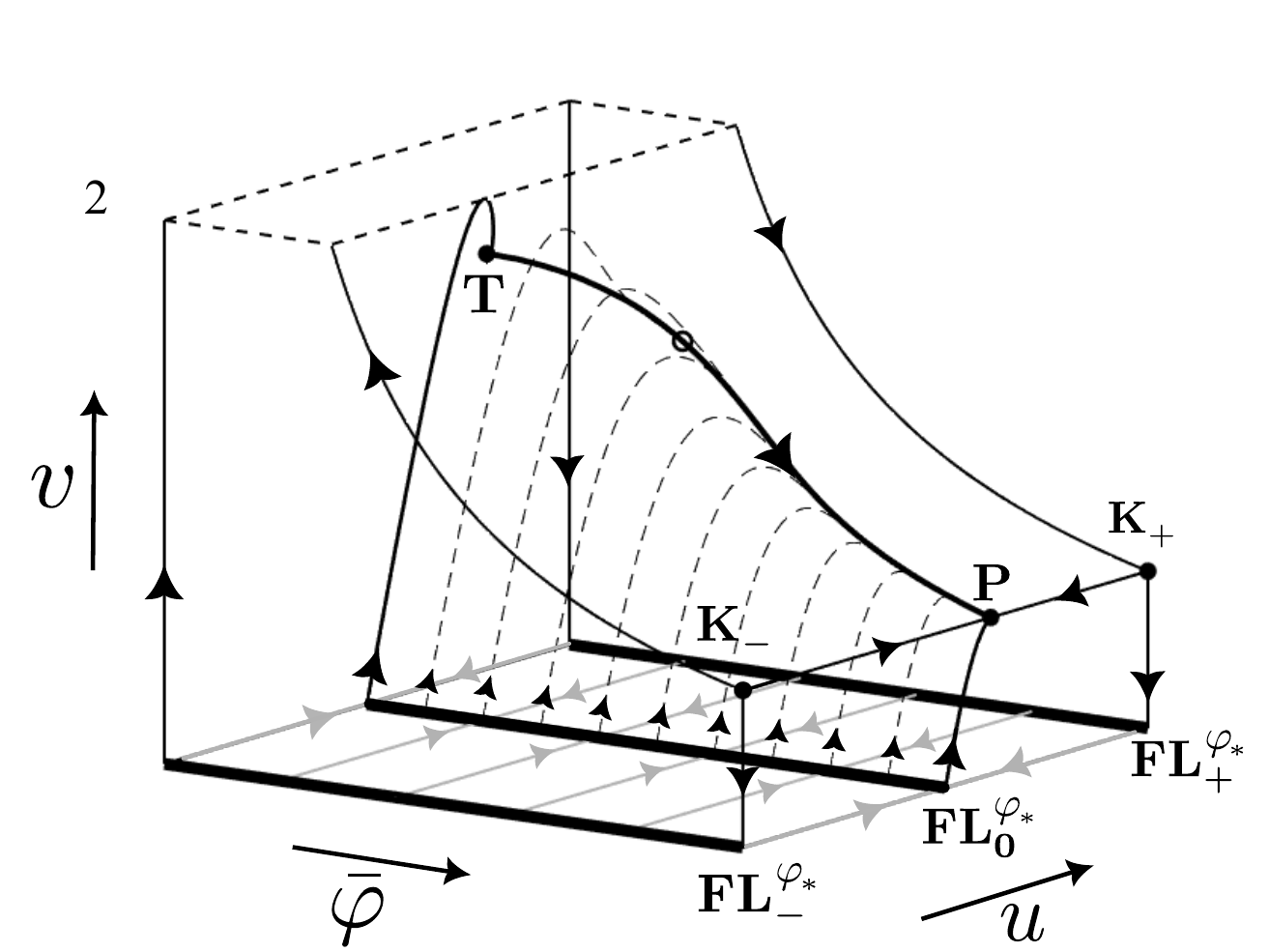}}
		\subfigure[$p=\frac12$, $\nu = 2\sqrt{\frac35}$]{\label{fig:track_const}
			\includegraphics[width=0.30\textwidth]{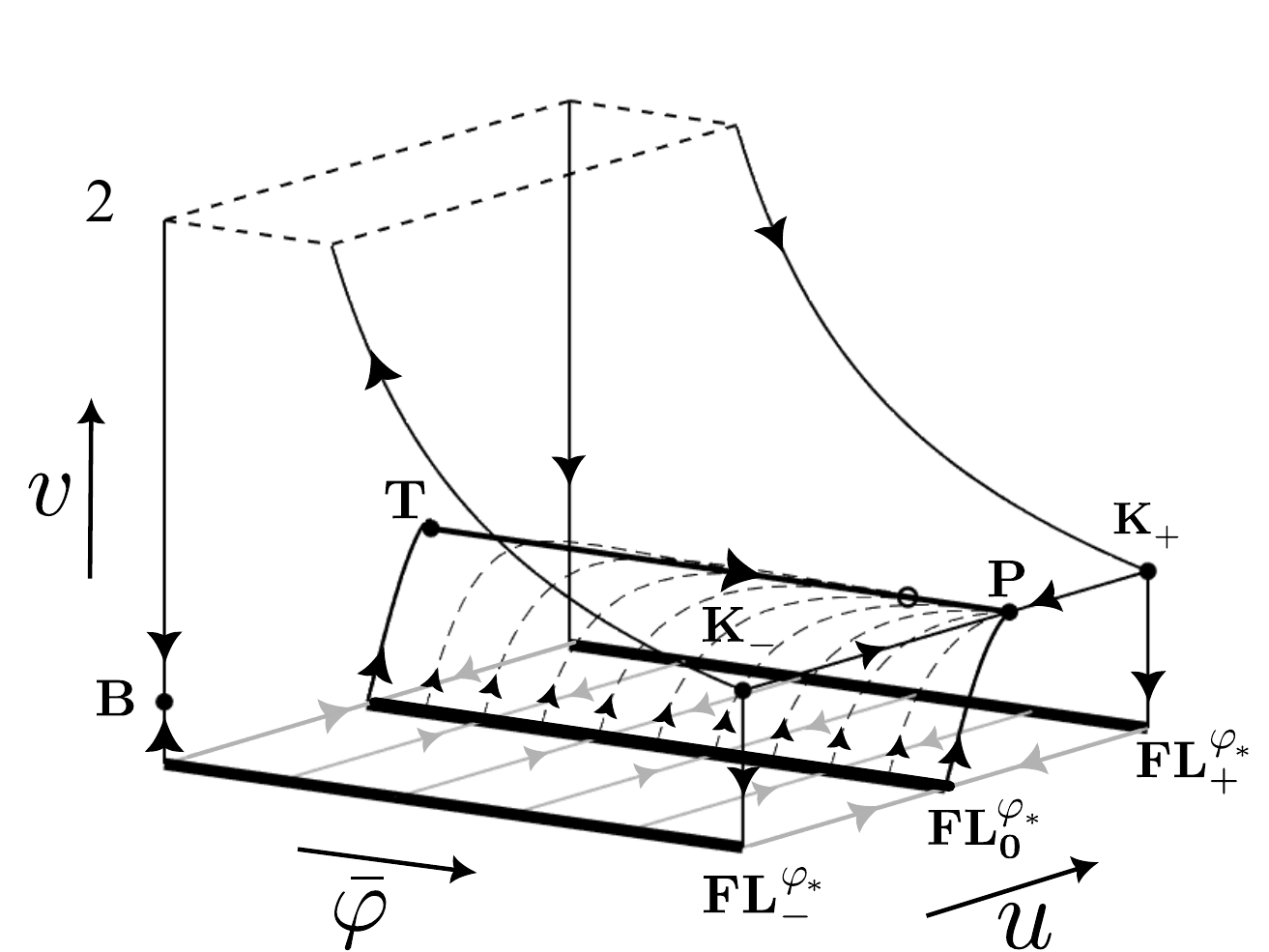}}
		\subfigure[$p=18$, $\nu = \frac{1}{6\sqrt{10}}$]{\label{fig:track_thaw}
			\includegraphics[width=0.30\textwidth]{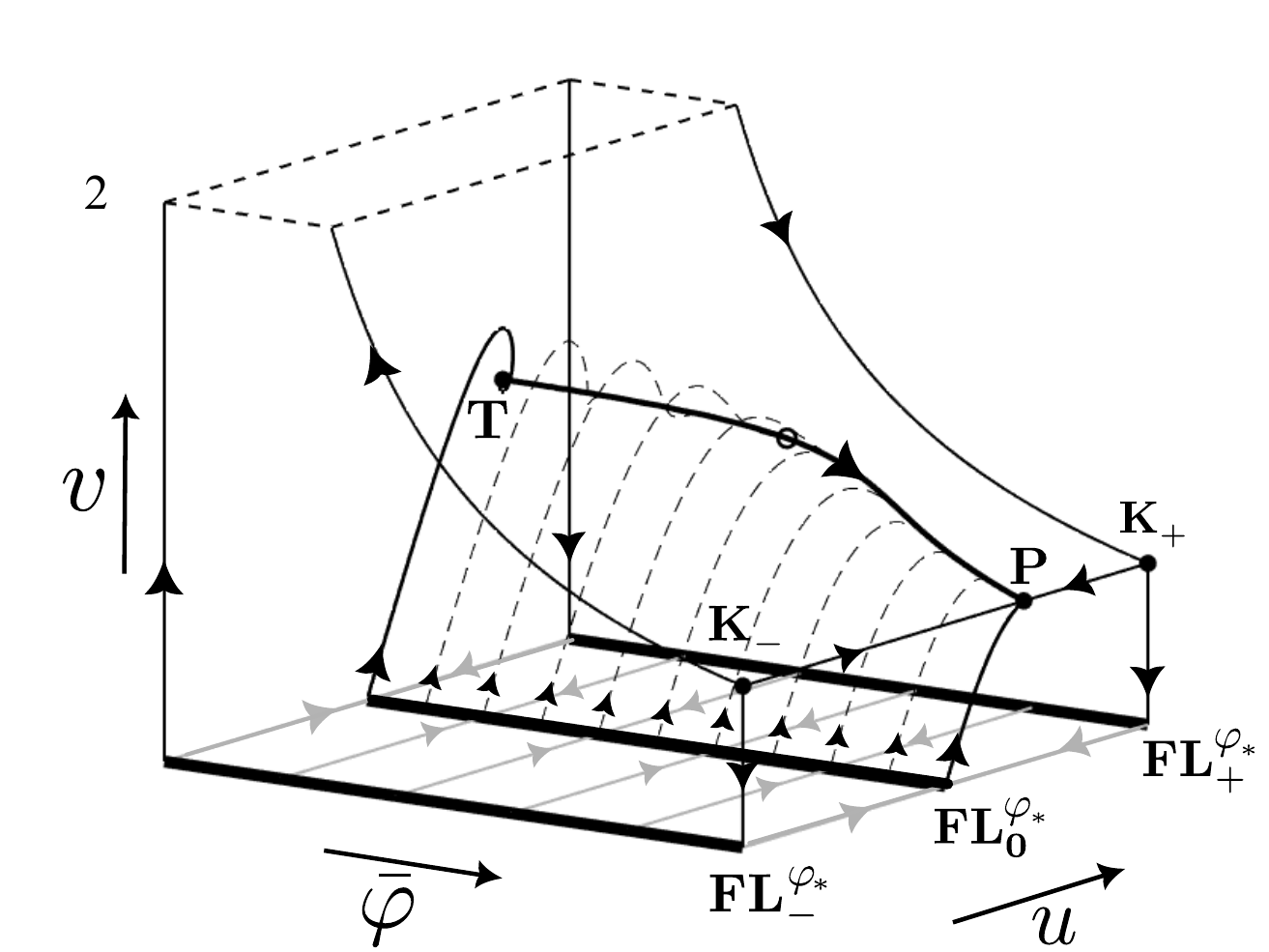}}
		\vspace{-0.5cm}
	\end{center}
	\caption{Illustration of the three cases of tracking quintessence attractor solutions
with (a) $w_\varphi|_\mathrm{T} = -0.7\,\rightarrow\, w_\varphi|_\mathrm{P} = -0.9$,
(b) $w_\varphi = w_\varphi|_\mathrm{T} = w_\varphi|_\mathrm{P} = -0.8$,
(c) $w_\varphi|_\mathrm{T} = -0.9\,\rightarrow\, w_\varphi|_\mathrm{P} = -0.7$,
for $V = V_*\sinh^{-p}(\nu\varphi)$ (the ring denotes when $\Omega_\varphi = 0.68$ for these 
solutions), and how they form the interior boundary of the thawing quintessence attractor 
solutions.
}
\label{fig:track}
\end{figure}

Since $u$, which we keep, is bounded, so is $v\lambda(\bar{\varphi})$ in
eq.~\eqref{u.prime}. To regularise the equations we absorb the asymptotic factor in
$\lambda(\bar{\varphi})$ into $v$, leading to the following generalised definition
of $v$:
\begin{equation}\label{vgen}
v \equiv \bar{\varphi}^{-k}\sqrt{\frac{\Omega_\varphi}{3}},
\end{equation}
where $k=0$ for the previous bounded $\lambda$ models whereas we set $k=1$ for the present
asymptotically inverse power-law potentials. Thus $v$
becomes unbounded when $\bar{\varphi}\rightarrow 0$, leading to the
\emph{`ski-slope' state space} in Figure~\ref{fig:track} with \emph{two}
matter-dominated boundaries ($\Omega_\mathrm{m}=1$), $v=0$, $\bar{\varphi}=0$,
and the regular dynamical system,
\begin{subequations}\label{Dynsysuvii}
\begin{align}
\bar{\varphi}^\prime &= 3\nu\bar{\varphi}(1-\bar{\varphi}^2)\,uv,\\
u^\prime &= \frac32(2-u^2)(p\nu\,v - u),\\
v^\prime &= \frac32\left[(1 - u^2)(1 - 3\bar{\varphi}^2v^2) - 2\nu(1-\bar{\varphi}^2)\,uv\right]v.
\end{align}
\end{subequations}

Although $v$ is unbounded all physical tracking quintessence features are
in a bounded part of the state space: the matter-dominated tracker fixed point
$\mathrm{T}$, which is a spiral sink \emph{on} the matter-dominated 
$\bar{\varphi} = 0$ boundary, is given by
$(\bar{\varphi},u,v) = u_\mathrm{T}(0, 1, 1/p\nu)$, $u_\mathrm{T} = \sqrt{\frac{p}{2+p}}>0$,
which together with that $u^\prime >0$ when $u\leq 0$ ensures that the tracking
quintessence attractor solution $\mathrm{T}\rightarrow \mathrm{P}$, the unstable manifold
of $\mathrm{T}$, is in a bounded part with $u>0$ of the interior state space,
with $\mathrm{P}$ $(\bar{\varphi},u,v) = (1,p\nu/\sqrt{3},1/\sqrt{3})$ being the future
global attractor. \emph{The tracking quintessence attractor mechanism}
is similar to the scaling freezing attractor mechanism, i.e., an open set
of solutions are pushed toward the `attractor' solution $\mathrm{T}\rightarrow \mathrm{P}$
near $\mathrm{T}$ by the unstable manifold of $\mathrm{T}$ (the $\bar{\varphi} = 0$ boundary
set), resulting in that these solutions `track', i.e. \emph{shadow},
$\mathrm{T}\rightarrow \mathrm{P}$, which thereby approximate them extremely well
during their observationally relevant evolution, and where, similarly to the scaling
freezing case, $\mathrm{T}\rightarrow \mathrm{P}$ is the interior boundary of
${\bf U}\mathrm{FL}_0^{\varphi_*}$.

We then note that
\begin{equation}
1 + w_\varphi|_\mathrm{T} = u_\mathrm{T}^2 = \frac{p}{2 + p},\qquad
1 + w_\varphi|_\mathrm{P} = u_\mathrm{P}^2 = \frac13\lambda_+^2 = \frac13(p\nu)^2,
\end{equation}
which results in three cases:
\begin{subequations}\label{w.tracker.2}\
\begin{alignat}{2}
u_\mathrm{T} &= \sqrt{\frac{p}{2+p}} > u_\mathrm{P} = p\nu/\sqrt{3} &\quad \Rightarrow\quad \nu &< \sqrt{\frac{3}{p(2+p)}},\\
u_\mathrm{T} &= \sqrt{\frac{p}{2+p}} = u_\mathrm{P} = p\nu/\sqrt{3} &\quad \Rightarrow\quad \nu &= \sqrt{\frac{3}{p(2+p)}},\\
u_\mathrm{T} &= \sqrt{\frac{p}{2+p}} < u_\mathrm{P} = p\nu/\sqrt{3} &\quad \Rightarrow\quad \nu &> \sqrt{\frac{3}{p(2+p)}},
\end{alignat}
\end{subequations}
where the second case leads to a tracking constant attractor
quintessence solution $\mathrm{T}\rightarrow \mathrm{P}$ that
is a straight line in the state space $(\bar{\varphi}, u,v)$ with
$u=u_\mathrm{T} = \sqrt{p/(2 + p)}$, $v = v_\mathrm{T} = u_\mathrm{T}/p\nu =1/\sqrt3$,
which yields $1+w_\varphi = u_\mathrm{T}^2 = p/(2+p)= \mathrm{const.}$ throughout the
entire evolution of the solution with
$\Omega_\varphi \equiv 3v^2\bar{\varphi}^2 = \bar{\varphi}^2$.\footnote{This
solution was found by Sahni and Starobinsky (2000)~\cite{sahsta00} and Urena-Lopez
and Matos (2000)~\cite{uremat00}.} In contrast, the first (last) case 
results in a tracking freezing (thawing)
quintessence attractor solution, but in all cases $\mathrm{T}\rightarrow \mathrm{P}$
is the interior boundary of the thawing quintessence surface of solutions,
${\bf U}\mathrm{FL}_0^{\varphi_*}$. Appropriate $p$ and $\nu$ yield
viable tracking quintessence attractor solutions and thereby also
viable thawing quintessence solutions, see Figure~\ref{fig:track}.

Next we consider the inverse power-law potential (Case I) and a generalisation of
$V \propto \exp(M/\varphi) - 1$ (Case II):
\begin{subequations}
\begin{alignat}{3}
V &= V_*\varphi^{-p},&\qquad p &> 0,&\qquad\lambda &= \frac{p}{\varphi},\\
V &= V_*\left[\exp\left(\frac{p}{r}\varphi^{-r}\right) - 1\right],&\qquad p &> 0,\quad r>0,&\qquad\lambda &= \frac{p\varphi^{-(1+r)}}{1 - \exp\left(-\frac{p}{r}\varphi^{-r}\right)}.
\end{alignat}
\end{subequations}
%
%
%
We choose
\begin{equation}
\bar{\varphi} \equiv \frac{\varphi^{\frac{1}{m}}}{1 + \varphi^{\frac{1}{m}}},\qquad \varphi = \left(\frac{\bar{\varphi}}{1 - \bar{\varphi}}\right)^m,\qquad
\text{with}\quad m=1\quad\text{in}\;\, \text{Case}\;\, \text{I}.
\end{equation}

\begin{figure}[ht!]
	\begin{center}
		\subfigure[$p=1$, $r=0$]{\label{fig:track_r0}
			\includegraphics[width=0.30\textwidth]{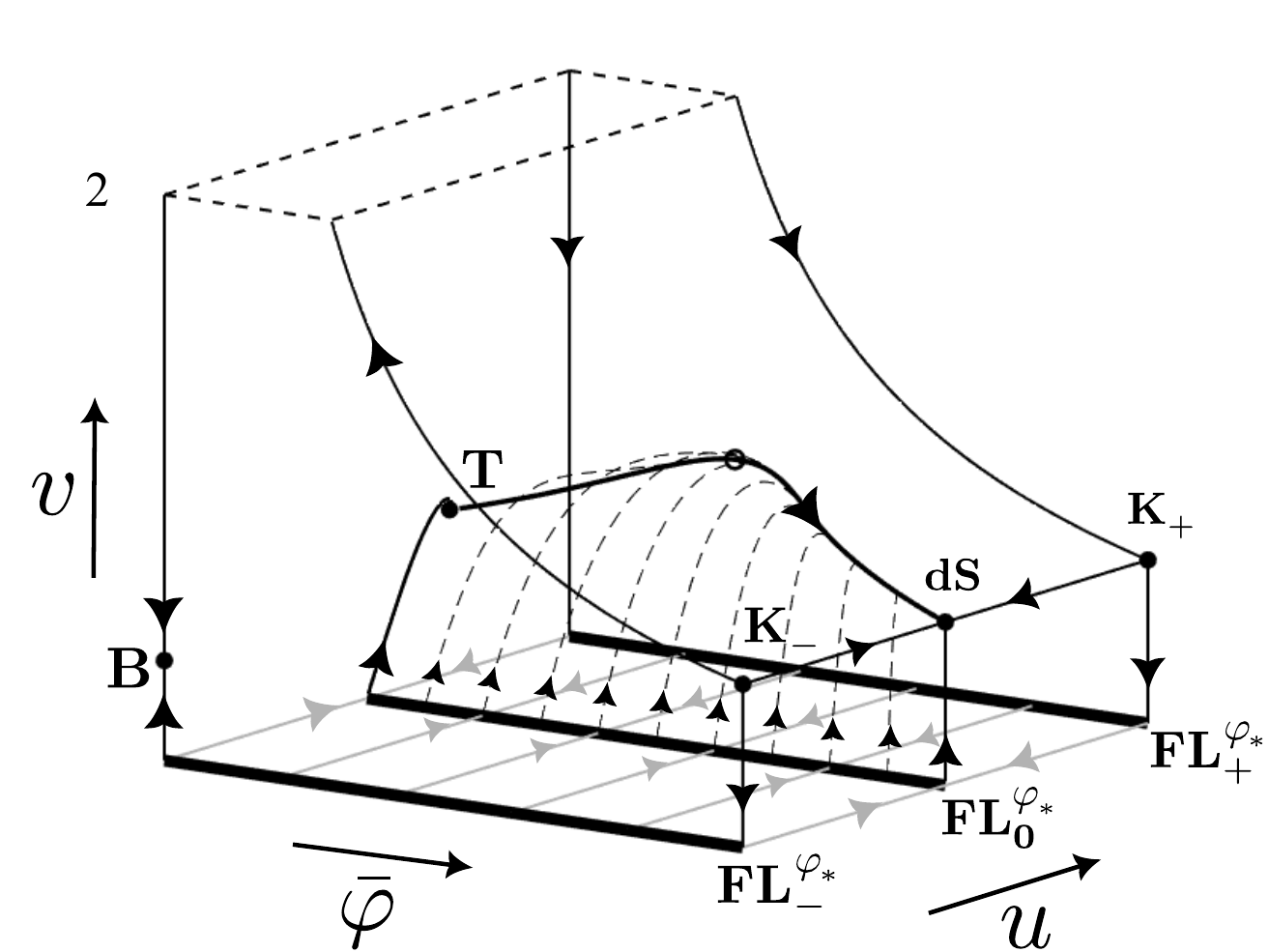}}
		\hspace{1.0cm}
		\subfigure[$p=1$, $r=1$]{\label{fig:track_r1}
			\includegraphics[width=0.30\textwidth]{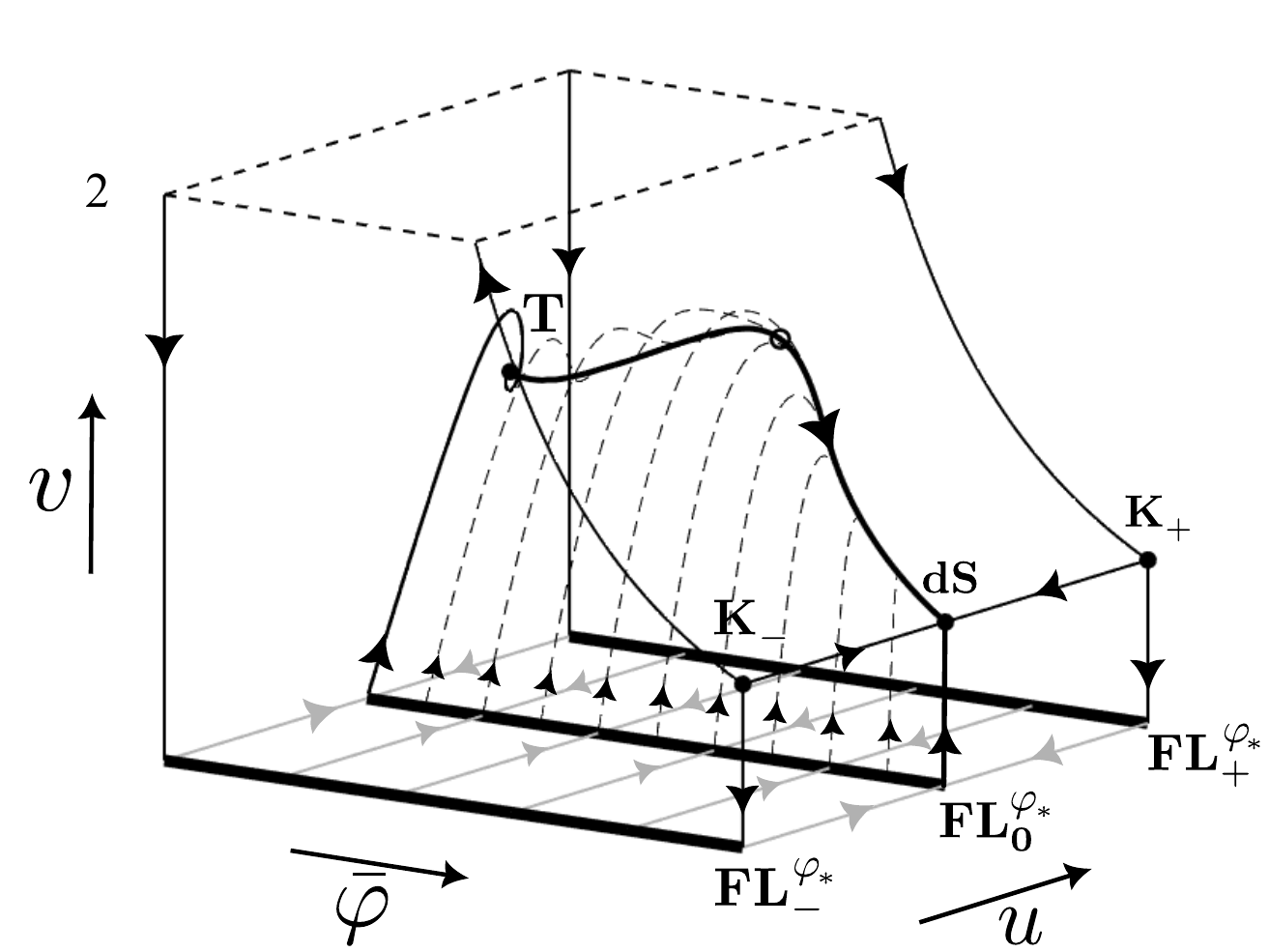}}
		\vspace{-0.5cm}
	\end{center}
	\caption{The figures depict the tracking attractor quintessence solution
$\mathrm{T}\rightarrow\mathrm{dS}$ (the ring denotes when $\Omega_\varphi = 0.68$),
showing that it is the interior boundary of the surface of thawing
quintessence solutions. The $\mathrm{FL}_0^{\varphi_*}\rightarrow\mathrm{T}$
solution on the boundary $\bar{\varphi}=0$ illustrates that the tracker fixed point $\mathrm{T}$
is a stable focus on this boundary; (a) shows the key quintessence aspects for
$V \propto \varphi^{-p}$ whereas (b) depicts them
for $V \propto \exp(\frac{p}{r}\varphi^{-r}) - 1$. Note that
the tracking quintessence solution $\mathrm{T}\rightarrow\mathrm{dS}$ in (b) is the unstable
\emph{centre} manifold of $\mathrm{T}$.
}
\label{Fig:trackinv}
\end{figure}

Choosing $k=1+r$ in the definition of $v$ in~\eqref{vgen}, with $r=0$ in Case I,
leads to
\begin{subequations}\label{DynsysuvI_II}
\begin{align}
\bar{\varphi}^\prime &= 3\bar{\varphi}\,\tilde{G}(\bar{\varphi})\,uv,\\
u^\prime &= \frac32(2-u^2)(v\,H(\bar{\varphi}) - u),\\
v^\prime &= \frac32\left[(1 - u^2)(1 - 3v^2\bar{\varphi}^{2(1+r)m}) - 2(1+r)m\,\tilde{G}(\bar{\varphi})\,uv\right]v,
\end{align}
\end{subequations}
where
\begin{subequations}
\begin{alignat}{5}
\mathrm{Case\, I}\!: &\quad \tilde{G} &=& (1 - \bar{\varphi})^2,&\qquad H &=& p(1- \bar{\varphi}),\hspace{2.3cm} &\qquad r\, &=&\, 0,\\
\mathrm{Case\, II}\!: &\quad \tilde{G} &=& \frac{1}{m}\bar{\varphi}^{rm}(1 - \bar{\varphi})^{1+m},&\qquad
H &=& \frac{p(1 - \bar{\varphi})^{(1+m)r}}{1 - \exp\left[-\frac{p}{r}\left(\frac{1 - \bar{\varphi}}{\bar{\varphi}}\right)^{rm}\right]},&\qquad r\, &>&\, 0.
\end{alignat}
\end{subequations}
This dynamical system is regular for Case I, for which $r=0$, $m=1$, with $\mathrm{T}$ located at
$(\bar{\varphi},u,v)=(0,\sqrt{p/(2+p)},1/\sqrt{p(2+p)})\,\Rightarrow\, w_\varphi = -2/(2+p)$.
The system is also regular in Case II when $r$ is a rational number and $m$ is chosen
as its denominator; for irrational values of $r$ one can obtain arbitrary differentiability
by choosing $m$ sufficiently large. In Case II $\mathrm{T}$ is located at
$(\bar{\varphi},u,v)=(0,1,1/p)\,\Rightarrow\,w_\varphi = 0$.
This might tempt one to conclude that the tracking quintessence solution is a
scaling freezing quintessence solution, but this is not the case since
$\lim_{N\rightarrow-\infty}\rho_\varphi/\rho_\mathrm{m} = 0$ and not a positive constant as it was
for scaling freezing quintessence. This is due to that the tracking quintessence solution now
is the unstable \emph{centre} manifold of $\mathrm{T}$ corresponding to an eigenvalue that is zero.
The (stronger) attracting properties of this solution is thereby more like the inflationary attractor
solutions than the previous scaling freezing and tracking quintessence models, which corresponded to
an unstable manifold associated with a positive eigenvalue, illustrated in
Figures~\ref{fig:alpha_attr} and~\ref{fig:track}, respectively.
One example of each of the present cases is given in Figure~\ref{Fig:trackinv}, showing that the
tracking quintessence attractor solution is the interior boundary of the
thawing quintessence ${\bf U}\mathrm{FL}_0^{\varphi_*}$ surface of solutions.
%
%

\section{Quintessence approximations\label{sec:approx}}

In~\cite{aluggla2025} (AU) new approximations were introduced for thawing
quintessence and for tracking quintessence with asymptotic inverse power-law
potentials. These approximations are simpler and more accurate than any previous
ones in the literature, as shown in~\cite{aluggla2025,shlivko2025}. They
are based on the DE assumptions: (i) $\lim_{N\rightarrow-\infty}w_\mathrm{DE} = w_\infty$,
$w_\infty \in [-1,0)$; (ii) past asymptotic matter-domination,
$\lim_{N\rightarrow-\infty}\Omega_\mathrm{m}=1$,
$\lim_{N\rightarrow-\infty}\Omega_\mathrm{DE}=0$. The field equations
for matter consisting of $\rho_\mathrm{m}$ and $\rho_\mathrm{DE}$ then results
in corrections by coupled series expansions in
\begin{equation}
T = T_0e^{-3w_\infty N},
\end{equation}
for $w_\mathrm{DE}$ and $\Omega_\mathrm{DE}$ given by
\begin{subequations}
\begin{align}
w_\mathrm{DE} &= w_\infty\left[1 - \alpha T + \alpha\beta T^2 + \dots\right], \label{wDEexp}\\
\Omega_\mathrm{DE} &= T\left(1 - (1 + \alpha)T + \left[1+\frac{1}{2}\alpha\left(4 + \alpha + \beta\right)\right]T^2 + \dots\right).\label{OphiDEexp}
\end{align}
\end{subequations}
The following Pad\'e approximant for $w_\mathrm{DE}$ is then constructed from~\eqref{wDEexp},
\begin{equation}
w_\mathrm{DE} = [1/1]_{w_\mathrm{DE}}(T) = 
w_\infty\left(1 - \frac{\alpha T}{1 + \beta T}\right)\quad \Rightarrow\quad
w_\mathrm{DE}^\prime = \frac{3w_\infty^2\alpha T}{(1+\beta T)^2}.\label{wDE11}
\end{equation}
Setting $\alpha=0$ leads to models with $w_\mathrm{DE} = w_\infty$ and thereby the
$\Lambda$CDM model when $w_\infty = -1$. If one is so inclined, it is possible to
solve for $\alpha$ and $\beta$ in terms of $w_{\mathrm{DE}0}$ and
$w_{\mathrm{DE}0}^\prime$ at a reference time
$t_0\,\Rightarrow\,N = \ln(a/a_0)=0$ (subscripts $_0$ refer to the reference
time $t_0$),\footnote{When $t_0$ refers to the present time it is easy to express results
in the redshift $z$ since then $\exp(-N) = 1 + z$.} which leads to
\begin{equation}
w_\mathrm{DE} - w_\infty = \frac{3w_\infty(w_{\mathrm{DE}0} - w_\infty)}{3w_\infty + \xi_0(1 - \exp(3w_\infty N))},\qquad
\xi \equiv \frac{(w_\mathrm{DE} - w_\infty)^\prime}{w_\mathrm{DE} - w_\infty},
\end{equation}
where $T_0$ drops out from the calculation (cf. with~\cite{shlivko2025} for the thawing case
$w_\infty = -1$). Note that although the parameter $w_\mathrm{DE}(N)$ is
not an observable quantity, it can, however, together with $\rho_\mathrm{m} \propto \exp(-3N)$, 
be used to compute observables such as the Hubble variable, see AU.
%
%
%
%

The importance of the $\mathrm{DE}$ results is that they yield simple and accurate
\emph{approximations} for thawing quintessence ($w_\infty = -1$)
and tracking quintessence for asymptotically inverse power-law potentials,
for which $w_\infty \in (-1,0)$, where $w_\mathrm{DE}$ in~\eqref{wDE11}
is replaced with
\begin{equation}
w_\varphi \approx w_\infty\left(1 - \frac{\alpha\,T}{1 + \beta T}\right).
\end{equation}
However, setting $\alpha$ and $\beta$, or, equivalently, $w_{\varphi 0}$ and $w_{\varphi 0}^\prime$,
to optimize quintessence approximations only amounts to curve fitting, see~\cite{shlivko2025}.
To obtain \emph{non-numerical predictive power}, equation~\eqref{OphiDEexp} was used in AU to
construct $\Omega_\varphi \approx [1/1]_{\Omega_\varphi}(T) = \frac{T}{1 + (\alpha + 1)T}$,
which results in\footnote{In AU  $\gamma = 1 + \alpha$ was used instead of $\alpha$ since this
leads to the simpler expression $T_0 = \Omega_{\varphi 0}/(1 - \gamma\Omega_{\varphi 0})$.}
\begin{equation}
T_0 = \frac{\Omega_{\varphi 0}}{\Omega_{\mathrm{m}0} - \alpha\Omega_{\varphi 0}}\qquad (\Omega_{\mathrm{m}0} + \Omega_{\varphi 0} = 1)
\end{equation}
and thereby
\begin{equation}
w_\varphi \approx [1/1]_{w_\varphi}(T) =
w_\infty\left(1 - \frac{\alpha\Omega_{\varphi 0}}{\beta\Omega_{\varphi 0} + (\Omega_{\mathrm{m}0} - \alpha\Omega_{\varphi 0})e^{3w_\infty N}}\right).
\end{equation}
%

As shown in AU, for thawing quintessence the field equations yield
\begin{equation}
\alpha = \left(\frac{2}{3}\right)^3\epsilon_*,\qquad \beta = \frac35 - \left(\frac{2}{3}\right)^3\epsilon_* + \frac{4}{15}\eta_*,
\end{equation}
where $\epsilon_*$ and $\eta_*$ are the slow-roll parameters
\begin{equation}\label{epseta}
\epsilon(\varphi) \equiv \frac{1}{2}\left(\frac{V_{,\varphi}}{V}\right)^2 = \frac{\lambda^2}{2},\qquad
\eta(\varphi) \equiv \frac{V_{,\varphi\varphi}}{V} = \lambda^2-\lambda_{,\varphi},
\end{equation}
computed at $\varphi = \varphi_*$ on $\mathrm{FL}^{\varphi_*}_0$ and thereby the frozen value
$\varphi  = \varphi_*(\bar{\varphi}_*)$ on the matter-dominated boundary $v=0$.
Tracking quintessence for asymptotic inverse power-law potentials yields more
complicated expressions involving
\begin{equation}\label{Gammadef}
\Gamma = \frac{V\,V_{,\varphi\varphi}}{V_{,\varphi}^2} = 1 + (\lambda^{-1})_{,\varphi};
\end{equation}
the parameters $\alpha$ and $\beta$ are given by
\begin{subequations}\label{alphabetatrac}
\begin{align}
\alpha &= - w_\infty^{-1}(1 - w_\infty^2)k,\label{alphatrack}\\
\beta &= \frac{2w_\infty^2(3w_\infty - 1) +
k\left(12w_\infty^4 - w_\infty^3 - 3w_\infty^2 + 2w_\infty-1\right) + k^{(2)}}{w_\infty(12w_\infty^2 - 3w_\infty +1)},\\
k &= \frac{w_\infty - \frac23\Gamma^{(1)}}{4w_\infty^2 - 2w_\infty + 1},\qquad
k^{(2)} = \frac{w_\infty\Gamma^{(2)}}{9(w_\infty + 1)k},
\end{align}
\end{subequations}
where
\begin{equation}\label{Gammaseries}
\Gamma(\tilde{\varphi}) \approx \Gamma^{(0)} + \Gamma^{(1)}\tilde{\varphi} + \frac12\Gamma^{(2)}\tilde{\varphi}^2 + \dots,\qquad
\Gamma^{(n)} = \left. \frac{d^n\Gamma}{d\tilde{\varphi}^n}\right|_{\tilde{\varphi}=0},\qquad \Gamma^{(0)} = 1 + p^{-1},
\end{equation}
and
%
$\tilde{\varphi} \equiv \lambda^{-2}(\varphi)$.
%
Note that for $V\propto \varphi^{-p}$ it follows that $\Gamma = 1 + p^{-1} = \mathrm{const.}$ and
$w_\infty = p/(2+p)$, which makes the above expressions collapse to relatively simple expressions.

\section{Discussion\label{sec:disc}}

The methods and results in this paper are easy to generalise, for both various DE models and potentials.
For example, the potential $V = V_*\left[\exp\left(\frac{p}{r}\varphi^{-r}\right) - 1\right]$
belongs to a much larger class of potentials that is straightforward to treat with
the present ideas and techniques:
\begin{equation}\label{Vgen}
V = V_*\exp[pf^{-r}(\varphi)/r],\qquad V = V_*\left(\exp[pf^{-r}(\varphi)/r] - 1)\right);\qquad
V_* > 0,\quad r>0,
\end{equation}
where $\lim_{\varphi\rightarrow 0}[\varphi^{-1}f(\varphi)] = 1$,
exemplified by (cf. eq.~\eqref{tanhpot}):
\begin{equation}
f = \frac{\sinh(\nu\varphi)}{\nu\cosh^{1-\alpha}(\nu\varphi)}
= \frac{\tanh(\nu\varphi)}{\nu[1 - \tanh^2(\nu\varphi)]^{\alpha/2}}.
\end{equation}
Choosing $u$ and $v$ as in Section~\ref{sec:track} and
\begin{equation}
\bar{\varphi} = \tanh^{\frac{1}{m}}(\nu\varphi) \in [0,1],\qquad m>0,\quad \nu>0,
\end{equation}
yields dynamical systems problems resembling those in the previous section.

We have only considered monotonically decreasing potentials, but
there are also, e.g., potentials with positive minima that are cosmologically
interesting, and with new features. For example, a potential with two exponential
terms, as in~\eqref{two.exp.pot}, with $\lambda_->\sqrt{6}$ and
$\lambda_+< - \sqrt{6}$ yields a potential with a positive minimum and a past
attractor that is \emph{not} a fixed point but a so-called heteroclinic cycle, see AUW1
for further details. Thus, local fixed point analysis does not always suffice
for a description of asymptotics and global structure, even for
quite simple models. In this context, it was only recently that
models with an exponential potential and matter with a linear equation of state
were analysed globally~\cite{alhetal22}, using the synergy between
Hamiltonian and dynamical systems formulations to derive monotonic functions,
which are crucial for global results. Initially these models
were analysed by using $\Sigma_\varphi=\varphi^\prime/\sqrt{6}$ and
$\sqrt{\Omega_\varphi}$ as variables, which subsequently in the literature 
have been used for a variety of more general potentials, even though they often 
are quite inappropriate, illustrating that some variables needlessly result
in nonregular equations or/and state space coordinate singularities leading
to unphysical fixed points, or sets of fixed points with overly many
eigenvalues with zero real parts.

Models with matter and a potential with zero minima or a minimum provide other
interesting examples, as illustrated by monomial potentials. For these models it is
better to use other variables than $(\bar{\varphi},\Sigma_\varphi,\sqrt{\Omega_\varphi})$
or $(\bar{\varphi},u,v)$, such as those in~\cite{alhugg15,alhetal15}, which result
in that the matter-dominated subset is described by an isolated $\mathrm{FL}$ fixed
point. In this case the one-parameter subset of thawing quintessence solutions
corresponds to the unstable manifold of $\mathrm{FL}$ with \emph{two}
positive eigenvalues while the stable subset, associated with a single negative
eigenvalue, pushes an open subset of solutions to this thawing attractor
quintessence solution subset. Note also that apart from the thawing quintessence
attractor solutions these models also admit two (equivalent) inflationary attractor
solutions. Furthermore, in~\cite{alhetal15} a new dynamical systems averaging method
was introduced to deal with the future asymptotics, which for some of these models
consist of periodic solution trajectories.

Even though two different dynamical systems formulations both yield regular and
bounded state spaces, one of these might still be preferable. For example, consider
the asymptotically inverse power-law potential problem for which we regularised
the equations by changing the definition of $v$ from $v = \sqrt{\Omega_\varphi/3}$
to $v = \bar{\varphi}^{-1}\sqrt{\Omega_\varphi/3}$. If we instead of changing $v$
choose a new time variable $\tilde{\tau}$ defined by
$d\tilde{\tau}/dN \equiv \bar{\varphi}^{-1}$ this also regularises the equations.
This, however, results in structures on the $\bar{\varphi}=0$ boundary that
prevent asymptotic analytic results and an identification of the past asymptotic
state of the tracking quintessence attractor solution.

Another example was given in AUW1 where two state space formulations
were used to deal with potentials with bounded $\lambda$:
$(\bar{\varphi},\Sigma_\varphi,\Omega_\mathrm{m})$ and
$(\bar{\varphi},u,v)$. Both yield regular equations and bounded state spaces.
The first formulation, however, collapses the matter-dominated boundary $v=0$
to a line of fixed points, which hides the evolution of $w_\varphi$ during
the matter-dominated epoch, easily seen in $u$, where $w_\varphi(N)$ is often
plotted in graphs in papers about quintessence.

Although having a global state space picture has several advantages, apart
from the most simple problems there typically are different physical and mathematical
structures at different parts of the state space. It is therefore often advantageous
to cover different parts of the state space with different variables
adapted to the local structures, e.g.,
the approximations in Section~\ref{sec:approx} were based on locally useful unbounded
scalar field variables adapted to the past asymptotic features of the thawing and tracking
quintessence attractor solutions.

In conclusion, a thorough understanding of a model, using dynamical systems reformulations
of the field equations as a powerful tool, requires variables that are adapted to the
physical and mathematical structures of the problem at hand.

Finally, note that some of the presently cited papers contain brief descriptions of the historical
developments concerning dynamical systems in cosmology with a focus on scalar field cosmology.
As described in~\cite{alhugg15}, this history begins in the mid 1980s with inflationary
models with potentials with zero minima. This was later followed by models with an exponential
potential, as described in~\cite{alhetal22}, but note also the book~\cite{col03} by A. Coley. 
See~\cite{alhugg23} and~\cite{alhetal24} for historical backgrounds on quintessential inflation 
and tracking quintessence, respectively. In addition, see the introductory survey by S. Cotsakis 
and A. Yefremov in this volume.

\subsection*{Acknowledgments}
A. A. is supported by FCT/Portugal through FCT/Tenure 2023.15700. TENURE.003/CP00055/CT00003, 
CAMGSD, IST-ID, grant No. UID/4459/2025, and projects 2024.04456.CERN, and  
H2020-MSCA-2022-SE EinsteinWaves, GA No. 101131233. A. A. would also like to thank the 
CMA-UBI in Covilh\~a for kind hospitality. C. U. would like to thank the CAMGSD, 
Instituto Superior T\'ecnico in Lisbon for kind hospitality and, finally, John Wainwright 
for friendship and several decades of collaboration, including on quintessence.

\bibliographystyle{unsrt}
\bibliography{../Bibtex/cos_pert_papers}

\begin{thebibliography}{10}

\bibitem{callin05}
R.~R. Caldwell and E.~V. Linder.
\newblock Limits of quintessence.
\newblock {\em Phys. Rev. Lett.}, {\bf 95}:141301, 2005.

\bibitem{alhetal23}
A.~Alho, C.~Uggla, and J.~Wainwright.
\newblock Quintessence from a state space perspective.
\newblock {\em Physics of the Dark Universe}, {\bf 39}:101146, 2023.

\bibitem{dimowe17}
K.~Dimopoulos and C.~Owen.
\newblock Quintessential inflation with $\alpha$-attractors.
\newblock {\em J. of Cosmology and Astroparticle Physics}, {\bf 06}:027, 2017.

\bibitem{alhugg23}
A.~Alho and C.~Uggla.
\newblock Quintessential $\alpha$-attractor inflation: A dynamical systems
  analysis.
\newblock {\em Journal of Cosmology and Astroparticle Physics}, {\bf 11}:083,
  2023.

\bibitem{akretal18}
Y.~Akrami et~al.
\newblock Dark energy, $\alpha$-attractors, and large-scale structure surveys.
\newblock {\em JCAP}, {\bf 06}:041, 2018.

\bibitem{steetal99}
P.~J. Steinhardt, L.~Wang, and I.~Zlatev.
\newblock Cosmological tracking solutions.
\newblock {\em Phys.\ Rev.\ D}, {\bf 59}:123504, 1999.

\bibitem{zlaetal99}
I.~Zlatev, L.~Wang, and P.~J. Steinhardt.
\newblock Quintessence, cosmic coincidence and the cosmological constant.
\newblock {\em Phys. Rev. Lett.}, {\bf 82}:896, 1999.

\bibitem{alhetal24}
A.~Alho, C.~Uggla, and J.~Wainwright.
\newblock Tracking quintessence.
\newblock {\em Physics of the Dark Universe}, {\bf 44}:101433, 2024.

\bibitem{tsu13}
S.~Tsujikawa.
\newblock Quintessence: a review.
\newblock {\em Class. Quantum Grav.}, {\bf 30}:214003, 2013.

\bibitem{sahsta00}
V.~Sahni and A.Starobinsky.
\newblock The case for a positive cosmological lambda-term.
\newblock {\em Int. J. Mod. Phys. D}, {\bf 9}:373, 2000.

\bibitem{uremat00}
L.A. Urena-Lopez and T.~Matos.
\newblock New cosmological tracker solution for quintessence.
\newblock {\em Phys. Rev. D}, {\bf 62}:081302, 2000.

\bibitem{aluggla2025}
Artur Alho and Claes Uggla.
\newblock New simple and accurate quintessence approximations.
\newblock {\em Phys. Rev. D}, 111:083549, Apr 2025.

\bibitem{shlivko2025}
David Shlivko, Paul~J. Steinhardt, and Charles~L. Steinhardt.
\newblock Optimal parameterizations for observational constraints on thawing
  dark energy.
\newblock {\em Journal of Cosmology and Astroparticle Physics}, 2025(06):054,
  jun 2025.

\bibitem{alhetal22}
A.~Alho, W.~C. Lim, and C.~Uggla.
\newblock Cosmological global dynamical systems analysis.
\newblock {\em Class. Quantum Grav.}, {\bf 39}:145010, 2022.

\bibitem{alhugg15}
A.~Alho and C.~Uggla.
\newblock Global dynamics and inflationary center manifold and slow-roll
  approximants.
\newblock {\em Journal of Mathematical Physics}, {\bf 56}(012502), 2015.

\bibitem{alhetal15}
A.~Alho, J.~Hell, and C.~Uggla.
\newblock Global dynamics and asymptotics for monomial scalar field potentials
  and perfect fluids.
\newblock {\em Class. Quant. Grav.}, {\bf 32}(14):145005, 2015.

\bibitem{col03}
A.~A. Coley.
\newblock {\em Dynamical systems and cosmology}.
\newblock Kluwer Academic Publishers, Dordrecht, 2003.

\end{thebibliography}

\end{document}